\documentclass[12pt]{iopart}
\usepackage{iopams}
\usepackage{graphicx}
\usepackage{pstricks}
\usepackage{pst-coil}
\usepackage{pst-node}
\usepackage{subfigure}

\begin{document}

\title{Cooling atoms into entangled states}
\author{Giovanni Vacanti and Almut Beige}
\address{The School of Physics and Astronomy, University of Leeds, Leeds LS2 9JT, United Kingdom}
\ead{a.beige@leeds.ac.uk}

\begin{abstract}
We discuss the possibility of preparing highly entangled states by simply cooling atoms into the ground state of an applied interaction Hamiltonian. As in laser sideband cooling, we take advantage of a relatively large detuning of the desired state, while all other qubit states experience resonant laser driving. Once spontaneous emission from excited atomic states prepares the system in its ground state, it remains there with a very high fidelity for a wide range of experimental parameters and all possible initial states. After presenting the general theory, we discuss concrete applications with one and two qubits.
\end{abstract}

\section{Introduction}

Dissipative systems are dynamical systems which lose energy over time. In classical physics, this typically happens due to friction or turbulence. In quantum mechanics, the loss of energy is due to processes like the spontaneous emission of photons. The observation of a photon as well as the observation of no photons reveals information about the system, thereby resulting in a so-called environment-induced measurement \cite{plenioknight}. These measurements can assist quantum computational tasks, like the controlled generation of entanglement, in many different ways \cite{Zurek,Cirac}. Measurement-based state preparation schemes are in general relatively simple and promise very high fidelities. In fact, their performance is widely independent of the concrete size of the experimental parameters. In general, it is only limited by the accuracy with which the relevant measurement outcomes can be detected. 

For example, no-photon measurements can improve the performance of adiabatic passages by stabilising the trajectory of the system from one quantum state into another \cite{Pellizzari1995,Marr}. They are also able to restrict the time evolution of a system onto a decoherence-free subspace, thereby resulting in very robust and relatively simple quantum computing schemes \cite{Beige,Dugic,Jcav,almutIontrap,Milburn}. Another way to take advantage of dissipation is to use photon measurements for the build up of highly entangled states for distributed quantum computing \cite{Cabrillo,Lim,Lim2,Browne}. The feasibility of this approach has already been demonstrated experimentally \cite{Grangier,Wilk,Monroe}. Moreover, continuous photon measurements can be used to control entanglement via direct quantum feedback \cite{wise,Metz,Metz2,busch,wise2}.

Alternatively, Verstraete {\em et al.}~\cite{Verstraete} recently suggested to engineer dissipative processes to prepare the ground states of frustration-free Hamiltonians without having to register measurement outcomes. These processes can be employed for efficient universal quantum computing. At the same time, Kraus {\em et al.}~\cite{Kraus} showed that dissipation can be used to prepare multipartite entangled states with efficient relaxation times in the number of qubits by designing the system interactions and environmental couplings such that the desired state is the stationary state of the system. In general, this requires the design of non-local jump operators. Only for certain target states, jump operators are found which need to act only on a few neighboring qubits. The scheme \cite{Kraus} moreover requires that the desired stationary state of the system is unique. If a system possesses more than one stationary state, small perturbations might result in sudden jumps between them. Examples are systems with macroscopic light and dark periods \cite{Dehmelt,Blatt}. These can be used to prepare highly entangled states upon the detection of a macroscopic dark period \cite{Metz,Metz2,busch}. See Ref.~\cite{viola} by Ticozzi and Viola for a general framework for the characterisation of attractive quantum Markovian dynamics and Refs.~\cite{schneider,squeeze,Diehl,Cho} for other related experimental proposals. 

In this paper we discuss how to cool qubits into the ground state of an applied interaction Hamiltonian $H_{\rm Int}$. The application of the proposed state preparation scheme to concrete physical systems of interacting atomic qubits, like atoms in optical lattices, ion traps, or atoms in optical tweezers, is straightforward, since we do not require complex system-reservoir interactions which result in non-local jump operators. Instead the cooling process is realised via laser driving of auxiliary atomic states and the free-space emission of photons. As we shall see below, our scheme is analog to laser sideband cooling. All qubit states other than the ground state of $H_{\rm Int}$ are resonantly driven by laser fields. When combined with dissipation, the result of this driving is the transfer of an arbitrary initial state into a highly entangled pure state \cite{Bartana}. Cooling atoms into entangled states promises high fidelities as long as the relevant coupling constants between qubits are much larger than the effective spontaneous decay rates of excited atomic states. 

There are five sections in this paper. In Section \ref{sec2} we introduce the qubit system and the cooling device considered throughout the paper. In Section \ref{sec3}, we describe the basic mechanism which transfers the qubits into the ground state of $H_{\rm Int}$. In Section \ref{sec4} we calculate the fidelity of the prepared state as a function of the system parameters and determine the corresponding cooling rates for concrete examples with one and two qubits. Finally, we summarise our results in Section \ref{sec6}.

\section{Theoretical model} \label{sec2}

In the following we consider a quantum system consisting of $N$ interacting atomic qubits and a laser cooling device. We introduce the notation that will be used throughout the paper and derive the Hamiltonian and the master equation of such a system.

\subsection{The qubits and their interaction}\label{SecHint}

Suppose $N$ qubits are encoded in the long living electronic ground states $|g_0\rangle \equiv |0 \rangle$ and $|g_1\rangle \equiv |1 \rangle$ of $N$ atoms in the presence of an interaction Hamiltonian $H_{\rm Int}$. The total Hamiltonian of the system reads in the Schr\"odinger picture
	\begin{eqnarray}\label{Hatoms} 
		H = H_{\rm Free} + H_{\rm Int} \, ,
	\end{eqnarray}  
where $H_{\rm Free}$ is the free Hamiltonian of the system. Here we denote the energy of the states $|g_0\rangle$ and $|g_1\rangle$ by $\hbar \omega_0$ and $\hbar \omega_1$. Hence
	\begin{eqnarray}
		H_{\rm Free} = \sum_{i=1}^N \sum_{j=0}^1 \hbar \omega_j \, |g_j \rangle_{ii} \langle g_j| \, .
	\end{eqnarray} 
Moreover, the interaction Hamiltonian $H_{\rm Int}$ can be written as 
\begin{eqnarray}\label{Hint}
		{H}_{\rm Int} &=& \sum_{n =0}^{2^N-1}  \lambda_n \, |\lambda_n \rangle \langle \lambda_n| \, ,
	\end{eqnarray}  
in terms of its eigenvectors and eigenvalues $|\lambda_n \rangle$ and $\lambda_n$.

A system with this Hamiltonian can be realised, for example, by trapping single atoms in the individual sites of an optical lattice. In this case, interactions between neighboring atoms can be induced by superposed laser fields which result in state-dependent tunneling rates and level shifts. Indeed it has been shown that it is possible to generate a variety of two and three-body interactions in this way \cite{plenio2}. The aim of this paper is to analyse a potential cooling mechanism to transfer the qubits into the ground state $|\lambda_0 \rangle$ of $H_{\rm Int}$. In case of a ground state degeneracy, the system is cooled into a mixed state.   

	\figure
	\quad\quad\quad\quad\quad 
	\psset{unit=0.8cm}
	\pspicture(13,13)\label{qubit}
	\rput{0}(4,1.5){\psline(0.5,1)(2.5,1)
	\psline(5,1)(7,1)
	\psline(0.5,6)(2.5,6)
	\psline(5,6)(7,6)
	\psline[linestyle=dashed](5,5)(7,5)
	\psline[linestyle=dashed](0.5,5)(2.5,5)
	\psline[linecolor=red]{<->}(1.5,1)(1.5,5)
	\psline[linecolor=red]{<->}(6,1)(6,5)
	\psline{<->}(0.7,5)(0.7,6)
	\psline{<->}(6.8,5)(6.8,6)
	
	\rput(2.1,3){ $\Omega_{0}^{(i,k)}$}
	\rput(5.3,3){$\Omega_{1}^{(i,k)}$}
	\rput(1.5,0.7){$|g_0\rangle=|0\rangle$}
	\rput(6,0.7){ $|g_1\rangle=|1\rangle$}
	\rput(1.5,6.3){ $|e_0\rangle$}
	\rput(6,6.3){ $|e_1\rangle$}
	\rput(0.4,3){$\Gamma_{00}$}
	\rput(7.1,3){$\Gamma_{11}$}
	\rput(2.8,4){$\Gamma_{01}$}
	\rput(4.8,4){$\Gamma_{10}$}
	\rput(0.3,5.5){$\Delta_k$}
	\rput(7.2,5.5){$\Delta_k$}
	\pszigzag[coilarm=.2,linearc=.1,coilwidth=.4]{->}(1,6)(1,1)
	\pszigzag[coilarm=.2,linearc=.1,coilwidth=.4]{->}(6.5,6)(6.5,1)
	\pszigzag[coilarm=.2,linearc=.1,coilwidth=.4]{->}(2,6)(5.5,1)
	\pszigzag[coilarm=.2,linearc=.1,coilwidth=.4]{->}(5.5,6)(2,1)
	\pscircle(3.8,3.2){4.6}}
	
	\pscurve{->}(2.5,10.2)(2.4,9)(2.8,8)(3.8,7)

	\rput{0}(0,-2){\pscircle*[fillcolor=black](1,14){0.3}
	\pscircle*[fillcolor=black](2.5,14){0.3}
	\pscircle*[fillcolor=black](4,14){0.3}
	\pscircle*[fillcolor=black](5.5,14){0.3}
	\pscircle*[fillcolor=black](6.5,14){0.05}
	\pscircle*[fillcolor=black](7,14){0.05}
	\pscircle*[fillcolor=black](7.5,14){0.05}
	\pscircle*[fillcolor=black](1,12.5){0.3}
	\pscircle*[fillcolor=black](2.5,12.5){0.3}
	\pscircle*[fillcolor=black](4,12.5){0.3}
	\pscircle*[fillcolor=black](5.5,12.5){0.3}
	\pscircle*[fillcolor=black](1,11.5){0.05}
	\pscircle*[fillcolor=black](1,11){0.05}
	\pscircle*[fillcolor=black](1,10.5){0.05}
	\pszigzag[coilarm=.2,linearc=.1,coilwidth=.4]{<->}(1.3,14)(2.2,14)
	\pszigzag[coilarm=.2,linearc=.1,coilwidth=.4]{<->}(2.8,14)(3.7,14)
	\pszigzag[coilarm=.2,linearc=.1,coilwidth=.4]{<->}(4.3,14)(5.2,14)
	\pszigzag[coilarm=.2,linearc=.1,coilwidth=.4]{<->}(1.3,12.5)(2.2,12.5)
	\pszigzag[coilarm=.2,linearc=.1,coilwidth=.4]{<->}(2.8,12.5)(3.7,12.5)
	\pszigzag[coilarm=.2,linearc=.1,coilwidth=.4]{<->}(4.3,12.5)(5.2,12.5)
	\pszigzag[coilarm=.2,linearc=.1,coilwidth=.4]{<->}(1,13.7)(1,12.8)
	\pszigzag[coilarm=.2,linearc=.1,coilwidth=.4]{<->}(2.5,13.7)(2.5,12.8)
	\pszigzag[coilarm=.2,linearc=.1,coilwidth=.4]{<->}(4,13.7)(4,12.8)
	\pszigzag[coilarm=.2,linearc=.1,coilwidth=.4]{<->}(5.5,13.7)(5.5,12.8)
	\pszigzag[coilarm=.2,linearc=.1,coilwidth=.4]{<->}(1,13.7)(1,12.8)}

	\rput(9,11.3){\large$H_{\rm Int}$}
	\pscurve{->}(8.4,11.3)(7,11.6)(5.8,11.3)
	\pscurve{->}(8.4,11.3)(7,10)(5,9.5)(4.8,10.2)
	\endpspicture
	
		\caption{The system consists of a collection of interacting atoms with ground states $|g_0\rangle$ and $|g_1\rangle$ which encode one qubit. Moreover, $K$ laser fields with different frequencies are applied which couple these states to the excited states $|e_0\rangle$ and $|e_1\rangle$ with the shown spontaneous decay rates. For simplicity we assume that the states $|g_j \rangle$ and $|e_j \rangle$, respectively, are of the same energy. Here $\Omega_j^{(i,k)}$ and $\Delta_k$ denote the Rabi frequency and the detuning of laser $k$ with respect to the $|g_j \rangle$ - $|e_j \rangle$ transition in atom~$i$.}	\label{device}
	\endfigure

\subsection{The cooling device} \label{SecCoolingDevice}

Our cooling device consists of a set of laser fields which couple the atomic ground states $|g_0 \rangle$ and $|g_1 \rangle$ to the auxiliary excited states $|e_0 \rangle$ and  $|e_1 \rangle$, respectively, as shown in \figurename~\ref{device}. We denote the energies of $|e_0 \rangle$ and  $|e_1 \rangle$ by $\hbar \tilde \omega_0$ and $\hbar \tilde \omega_1$. Since we added two levels to each atom,  the Hilbert space of the system is now of dimension $4^N$.  In the presence of the laser driving, the Hamiltonian in Eq.~(\ref{Hatoms}) hence becomes
	\begin{equation}\label{Hatoms2}
		H = H_{\rm Free} + H_{\rm Int} + H_{\rm Laser}
	\end{equation}
with $H_{\rm Free}$ now given by  
	\begin{eqnarray} \label{HFree}
		{H}_{\rm Free} &=& \sum_{i = 1}^N \sum_{j=0}^1 \hbar \omega_j \, |g_j \rangle_{ii} \langle g_j | + \hbar \tilde \omega_j \, |e_j \rangle_{ii} \langle e_j | \, .
	\end{eqnarray} 
Notice that the interaction Hamiltonian $H_{\rm Int}$ can no longer be written as in Eq.~(\ref{Hint}). In addition to the interactions between qubit states, couplings between states with at least one atom in the excited state have to be taken into account. 

To do so, we introduce additional states $|\lambda_n \rangle$ with $n$ ranging from $2^N$ to $4^N-1$. Suppose, these states form an orthonormal basis together with the $2^N$ qubit ground states $|\lambda_n \rangle$ in Eq.~(\ref{Hint}). Multiplying $H_{\rm Int}$ with the identities $\sum_{m=0}^{4^N-1}  |\lambda_m \rangle \langle \lambda_m|$ and $\sum_{n=0}^{4^N-1}  |\lambda_n \rangle \langle \lambda_n|$ from the left and from the right, respectively, we then find that it can be written as
	\begin{eqnarray}\label{Hint2}
		{H}_{\rm Int} &=& \sum_{n =0}^{2^N-1}  \lambda_n \, |\lambda_n \rangle \langle \lambda_n| \, + \sum_{m =2^N}^{4^N-1} \sum_{n =2^N}^{4^N-1} \tilde \chi_{mn} \, |\lambda_m \rangle \langle \lambda_n| \, .
\end{eqnarray}  
For reasons which become obvious later we do not assume that the newly introduced states $|\lambda_n \rangle$ with $n \ge 2^N$ are eigenvectors of $H_{\rm Int}$. The second term in  Eq.~(\ref{Hint2}) therefore contains non-diagonal terms with the $\tilde \chi_{mn} \equiv \langle \lambda_m | H_{\rm Int} | \lambda_n \rangle$ being coupling coefficients and diagonal terms with the $\tilde \chi_{nn} \equiv \langle \lambda_n | H_{\rm Int} | \lambda_n \rangle$ describing energy shifts.

Suppose the state $|g_0\rangle$ and $|g_1\rangle$ and the states $|e_0\rangle$ and $|e_1\rangle$, respectively, are of the same energy, {\em i.e.}
\begin{eqnarray} \label{long}
\omega_0 = \omega_1 \equiv \omega ~~ {\rm and} ~~ \tilde{\omega}_0 = \tilde{\omega}_1 \equiv \tilde \omega \, .
\end{eqnarray}
In this case, the detunings $\Delta_k$ of the applied laser fields depend neither on $i$ nor $j$. In the following, we denote the Rabi frequency of laser $k$ with respect to the $|g_j \rangle$ - $|e_j \rangle$ transition in atom $i$ by $\Omega^{(i,k)}_j$. If $K$ different laser fields are applied, then ${H}_{\rm Laser}$ equals in the usual rotating wave approximation 
	\begin{eqnarray} \label{Hlaser}
		{H}_{\rm Laser} &=& \sum_{i = 1}^N \sum_{j=0}^1 \sum_{k=1}^K {1\over 2} \hbar \Omega^{(i,k)}_j \, {\rm e}^{{\rm i} \big( \tilde \omega - \omega - \Delta_k \big) t} \, |e_j 		\rangle_{ii} \langle g_j | + {\rm H. c.} 
	\end{eqnarray}  
In general it is not possible to find an interaction picture which removes the time dependence from this Hamiltonian. 

\subsection{The effect of spontaneous emission} \label{reset}

In the following, we use the master equation to model spontaneous emission from the excited atomic states $|e_j \rangle$. If $\Gamma_j$ denotes the corresponding decay rate, then the density matrix of the atoms evolves according to 
	\begin{eqnarray} \label{mastereq}
		\dot \rho &=& - {{\rm i} \over \hbar} \, \big[ H , \rho \big] + {\cal L}(\rho)
	\end{eqnarray}
where the Lindblad operator ${\cal L}$ is given by
	\begin{eqnarray} \label{Lindblad}
		 {\cal L}(\rho) &=& \sum_{i=1}^N \sum_{j,k=0}^1 \Gamma_{jk} \, \Big[ \, {R}_{jk}^{(i)} \, \rho \, {R}_{jk}^{(i) \, \dagger} 
		 - {1 \over 2} \, R_{jk}^{(i) \, \dagger} R_{jk}^{(i)} \, \rho  - {1 \over 2} \, \rho \, R_{jk}^{(i) \, \dagger} R_{jk}^{(i)}  \, \Big] \, .
	\end{eqnarray}
The reset operators ${R}_{jk}^{(i)}$, defined as
	\begin{eqnarray} \label{resset}
		{R}_{jk}^{(i)} &\equiv & |g_j \rangle_{ii} \langle e_k| \, ,
	\end{eqnarray}
model transitions from $|e_k \rangle_i$ into $|g_j \rangle_i$ with $\Gamma_{jk}$ being the respective spontaneous decay rate. The overall decay rate of $|e_k \rangle_i$ equals $\Gamma_j = \Gamma_{j0} + \Gamma_{j1}$.

Although the reset operators in Eq.~(\ref{resset}) are local, the state of the atoms immediately after an emission can be highly entangled \cite{Schon}. The reason for this is that the state of a single system after a photon emission is a function of the direction of the emitted photon. In other words, the state of the atoms immediately after an emission can be almost any state. In many cases, the reset state has some overlap with the state $|\lambda_0 \rangle$ which we want to prepare.  As we see in the following, it is the interplay between the interaction Hamiltonian $H_{\rm Int}$ and spontaneous emission which is responsible for the eventual cooling of the atoms into $|\lambda_0 \rangle$. 

\section{The cooling mechanism} \label{sec3}

The state preparation scheme proposed in this paper is analogous to the well-established technique of laser sideband cooling. Both schemes require non-resonant laser driving and the spontaneous emission of photons. In both cases, the target state is one which is easy to reach but difficult to leave. After a certain transition time, one can therefore be sure that the system is with a very high fidelity in this state. Since the idea of state preparation via cooling can be understood easily by comparison, we start this section with a short overview on laser sideband cooling.  We then refer to the one-qubit case and consider a generalisation of the proposed state preparation scheme to $N$ qubits.

\subsection{Laser sideband cooling of a single atom}

	\figure
	\quad\quad\quad\quad\quad
	\psset{unit=1.2cm}
	\pspicture(8,8)\label{qubit}
	\rput(1.4,-1){
	\psline(-1,2)(6,2)
	\psline(-1,2.5)(6,2.5)
	\psline(-1,3)(6,3)
	\psline(-1,6.5)(6,6.5)
	\psline(-1,7)(6,7)
	\psline(-1,7.5)(6,7.5)
	\psline[linestyle=dashed](0,6)(1,6)
	\psline[linecolor=red]{<->}(0.5,2)(0.5,6)
	\psline[linecolor=red]{<->}(1.3,2.5)(1.3,6.5)
	\psline[linecolor=red]{<->}(2.1,3)(2.1,7)
	\pszigzag[coilarm=.2,linearc=.1,coilwidth=.4]{->}(5.1,7.5)(5.1,3)
	\pszigzag[coilarm=.2,linearc=.1,coilwidth=.4]{->}(4.3,7)(4.3,2.5)
	\pszigzag[coilarm=.2,linearc=.1,coilwidth=.4]{->}(3.5,6.5)(3.5,2)
	\pscircle*[fillcolor=black](-0.6,3.4){0.02}
	\pscircle*[fillcolor=black](-0.6,3.2){0.02}
	\pscircle*[fillcolor=black](-0.6,3.6){0.02}
	\pscircle*[fillcolor=black](-0.6,7.9){0.02}
	\pscircle*[fillcolor=black](-0.6,7.7){0.02}
	\pscircle*[fillcolor=black](-0.6,8.1){0.02}
	\rput(6.7,6.5){ $|2,m=0\rangle$}
	\rput(6.7,7){ $|2,m=1\rangle$}
	\rput(6.7,7.5){ $|2,m=2\rangle$}
	\rput(0,-4.5){
	\rput(6.7,6.5){ $|1,m=0\rangle$}
	\rput(6.7,7){ $|1,m=1\rangle$}
	\rput(6.7,7.5){ $|1,m=2\rangle$}}
	\psline{<->}(0.2,6)(0.2,6.5)
	\rput(-0.2,6.25){$\nu$}
	\rput(0.17,4){$\Omega$}
	\rput(0.97,4.5){$\eta \Omega$}
	\rput(1.77,5){$\eta \Omega$}
	\rput(3.2,4){$\Gamma$}
	\rput(4,4.5){$\Gamma$}
	\rput(4.8,5){$\Gamma$}}
	\endpspicture
		\caption{Level scheme for sideband cooling of a single atom which is based on the resonant driving of the $|1, m \rangle$ - $|2, m-1 \rangle$ transitions, while all other transitions are detuned by the phonon frequency $\nu$. The ground state $|1,m=0 \rangle$ is therefore the only off-resonant state. Crucial for the cooling to work is moreover spontaneous emission of photons from level 2 which preserves the phonon number $m$.}\label{sideband}
	\endfigure

Sideband cooling is an experimental technique commonly used to transfer single trapped atoms and ions close to absolute zero temperature \cite{Wineland,Eschner}. This is achieved with the help of an appropriately detuned laser field which couples the electronic states $|1 \rangle$ and $|2 \rangle$ with coupling strength $\eta \Omega$ to their quantised motion. For simplicity we consider only a one-dimensional trapping potential and denote its phonon states by $|m \rangle$. If the laser is red-detuned and its detuning $\Delta$ equals the phonon frequency $\nu$, then atomic transitions which reduce $m$ by one are in resonance, as shown in Figure \ref{sideband}. Simultaneously, unwanted transitions which increase $m$ or keep it constant are out of resonance. Suppose, the atom is initially in its ground state and has exactly $m$ phonons. The laser then couples this $|1, m \rangle$ state to $|2, m-1 \rangle$ and $|2, m \rangle$ with the transition into $|2, m- 1 \rangle$ being the most effective. If the atom now emits a photon, its state changes into $|1, m-1 \rangle$ with the population in $|1, m\rangle$ being almost negligible. Compared to the initial state, one phonon is permanently lost.

Efficient sideband cooling requires that the laser detuning, i.e.~the phonon frequency $\nu$, is much larger than the laser Rabi frequency $\Omega$ and the decay rate $\Gamma$ of level 2,
\begin{eqnarray} \label{comparison}
\nu &\gg & \Omega ~~ {\rm and} ~~ \Gamma \,. 
\end{eqnarray}
In this case, transitions out of the zero-phonon state $|1,0 \rangle$ which is the only non-resonantly driven state are very unlikely. Under ideal conditions, this is the final state prepared in the cooling process. Experiments show that its fidelity can be well above $99 \, \%$ \cite{Wineland,Eschner}. Conditions analog to Eq.~(\ref{comparison}) are in the following imposed on the proposed scheme for cooling atoms into an entangled state.

\subsection{1-qubit case}\label{1qubit}

In the case of only a single qubit, the preparation of the ground state $|\lambda_0 \rangle$ of $H_{\rm Int}$ requires only a single laser field with detuning $\Delta$. The relevant four-level scheme is shown in Figure \ref{device}. For simplicity we assume in the following that the (real) laser Rabi frequencies $\Omega_j^{(1)}$ and the spontaneous decay rates $\Gamma_{jk}$ are the same for both transitions, i.e.
	\begin{eqnarray}\label{CondParam}
		\Omega_j^{(1)} \equiv \Omega 
		~~ {\rm and} ~~ \Gamma_{jk} \equiv {1 \over 2} \Gamma \, .
	\end{eqnarray}
Taking this into account and transferring the Hamiltonian (\ref{Hatoms2}) into the interaction picture with respect to $H_0=H_{\rm Free} - \hbar \Delta (|\e_0 \rangle \langle e_0| + |\e_1 \rangle \langle e_1|)$ we obtain the time-independent Hamiltonian 
	\begin{eqnarray} \label{U0HlaserU0}
		H_{\rm I} &=& \sum_{j = 0}^1 {1 \over 2} \hbar \Omega \, |e_j \rangle \langle g_j | + {\rm H. c.} + \hbar \Delta |e_j \rangle \langle e_j | + H_{\rm Int} \, . 
	\end{eqnarray}
To transfer this Hamiltonian into a more useful form, we introduce the excited atomic states $|\lambda_2 \rangle$ and $|\lambda_3 \rangle$  as
	\begin{eqnarray} \label{sirene}
		|\lambda_{n+2} \rangle &\equiv & \left[ \sum_{j = 0}^1 |e_j \rangle \langle g_j | \right] | \lambda_n \rangle \, .
	\end{eqnarray}
Up to normalisation, these are the states $H_{\rm Laser} \, |\lambda_0 \rangle$ and $H_{\rm Laser} \, |\lambda_1 \rangle$ which couple to the eigenstates $|\lambda_0 \rangle$ and $|\lambda_1 \rangle$ of $H_{\rm Int}$ via laser driving. Notice that the states $|\lambda_2 \rangle$ and $|\lambda_3 \rangle$ are of the same energy as long as $H_{\rm Int}$ has no effect on the excited atomic states $|e_0 \rangle$ and $|e_1 \rangle$. Since also the ground states $|g_0 \rangle$ and $|g_1 \rangle$ are of the same energy (cf.~Eq.~(\ref{long})), the Hamiltonian (\ref{U0HlaserU0}) of the system in the interaction picture can be written in terms of the $|\lambda_n\rangle$ states as
	\begin{eqnarray}\label{H1}
		H_{\rm I} &=& \sum_{n=0,1} \frac{1}{2} \hbar \Omega \, |\lambda_n \rangle \langle \lambda_{n+2}|  + {\rm H.c.} +  \hbar (\lambda_n - \Delta) \, |\lambda_n \rangle \langle 				\lambda_n|
	\end{eqnarray}
up to an overall level shift $\hbar \Delta$ with no physical consequences. Moreover, the Lindblad operator in Eq.~(\ref{Lindblad}) now becomes
	\begin{eqnarray} \label{mastereq2}
		 {\cal L}(\rho) &=& \sum_{j=0,1} \sum_{k=2,3} {1 \over 2} \Gamma \, \Bigg[ \, \tilde {R}_{jk}\, \rho \, \tilde {R}_{jk}^{\dagger} 
		 - {1 \over 2} \, \tilde R_{jk}^{\dagger} \tilde R_{jk} \, \rho  - {1 \over 2} \, \rho \, \tilde R_{jk}^{\dagger} \tilde  R_{jk}  \, \Bigg] 
	\end{eqnarray}
with the new reset operators
	\begin{eqnarray}
		\tilde {R}_{jk} &\equiv & |\lambda_j \rangle \langle \lambda_k| \, . 
	\end{eqnarray}
In the derivation of this equations we took advantage of Eq.~(\ref{CondParam}) which assumes equal spontaneous decay rates for all transitions. 

	\figure
	\quad\quad\quad\quad\quad
	\psset{unit=1.2cm}
	\pspicture(8,8)
	\rput(1.4,-1){
	\psline(-1,2)(6,2)
	\psline(-1,3)(6,3)
	\psline(-1,7)(6,7)
	\psline[linestyle=dashed](0,6)(1,6)
	\psline[linecolor=red]{<->}(0.5,2)(0.5,6)
	\psline[linecolor=red]{<->}(1.5,3)(1.5,7)
	\pszigzag[coilarm=.2,linearc=.1,coilwidth=.4]{->}(4.5,7)(4.5,3)
	\pszigzag[coilarm=.2,linearc=.1,coilwidth=.4]{->}(3.5,7)(3.5,2)
	\rput(6.8,7){$|\lambda_2\rangle$, $|\lambda_3\rangle$}
	\rput(0,-4.5){
	\rput(6.5,6.5){ $|\lambda_0\rangle$}
	\rput(6.5,7.5){ $|\lambda_1\rangle$}}
	\psline{<->}(0.2,6)(0.2,7)
	\rput(-0.1,6.5){$\Delta_\lambda$}
	\rput(0.3,4.3){$\Omega$}
	\rput(1.3,4.8){$\Omega$}
	\rput(3.2,4.3){$\Gamma$}
	\rput(4.2,4.8){$\Gamma$}}
	\endpspicture
		\caption{Level scheme for cooling of a single atom into the eigenstate $|\lambda_0 \rangle$ of $H_{\rm Int}$. The laser drives only the $|\lambda_0 \rangle$ - $|\lambda_2 \rangle$ transition with detuning $\Delta_\lambda$ and the $|\lambda_1 \rangle$ - $|\lambda_3 \rangle$ transitions with zero detuning. The excited states $|\lambda_2 \rangle$ and $|\lambda_3 \rangle$ can decay into $|\lambda_2 \rangle$ and $|\lambda_3 \rangle$.}\label{qubit}
	\endfigure

Suppose the frequency of the applied laser field equals $\tilde \omega - \omega - \lambda_1$ which implies $\Delta = \lambda_1$. Then the ground state $|\lambda_1 \rangle$ is resonantly driven, while $|\lambda_0 \rangle$ experiences the detuning 
\begin{eqnarray} \label{Delta}
\Delta_\lambda &\equiv & \lambda_0 - \lambda_1\, ,
\end{eqnarray}
as shown in Fig.~\ref{qubit}. If $\Delta_\lambda$ is large compared to $\Omega$ and $\Gamma$, then the comparison with laser sideband cooling (cf.~Eq.~(\ref{comparison})) suggests that the system reaches $|\lambda_0 \rangle$ after a certain transition time with a very high fidelity. Notice that the detuning $\Delta_\lambda$ of $|\lambda_0 \rangle$ which we need to prepare the target state comes exactly from the fact that $|\lambda_0 \rangle$ is the ground state of the system.

\subsection{Generalisation to $N$ qubits} \label{moregeneral}

Let us now have a closer look at the case of $N$ laser-driven atomic qubits. For $N$ atoms, the relevant state space is of dimension $4^N$. Our task now consists of finding laser fields which couple the eigenstates $|\lambda_n \rangle$ of $H_{\rm Int}$ with $n$ between 1 and $2^N-1$ resonantly to excited atomic states while $|\lambda_0 \rangle$ remains off-resonance. Achieving this might require up to $2^N-1$ laser fields since there are $2^N$ atomic ground states. Choosing the right laser frequencies requires a detailed knowledge of the structure of $H_{\rm Int}$, since this Hamiltonian acts also on states with one atom in $|e_0 \rangle$ or $|e_1 \rangle$ and causes level shifts and interactions among them. Here we do not need to consider states with more than one atom in an excited state as long as these have sufficiently large spontaneous decay rates. As already mentioned above, the emission of a photon in general transfers the atoms into a states which has some overlap with $|\lambda_0 \rangle$.
  
	\figure
	\quad\quad\quad\quad\quad
	\psset{unit=1.2cm}
	\pspicture(8,8)
	\rput(1.4,-1)
	{
		\rput(0,-0.5)
		{
			\psline(-1,2)(6,2)
			\psline(-1,2.5)(6,2.5)
			\psline(-1,3)(6,3)
			\psline(-1,4.1)(6,4.1)
			\pscircle*[fillcolor=black](-0.6,3.2){0.02}
			\pscircle*[fillcolor=black](-0.6,3.4){0.02}
			\pscircle*[fillcolor=black](-0.6,3.6){0.02}
		}
		\rput(0,-5)
		{
			\rput(6.25,6.5){\footnotesize$|\lambda_0\rangle$}
			\rput(6.25,7){\footnotesize $|\lambda_1\rangle$}
			\rput(6.25,7.5){\footnotesize $|\lambda_2\rangle$}
			\rput(6.5,8.6){\footnotesize $|\lambda_{2^N-1}\rangle$}
		}
		\psline(-1,6.5)(6,6.5)
		\psline(-1,6.8)(6,6.8)
		\psline(-1,7.1)(6,7.1)
		\psline(-1,7.9)(6,7.9)
		\pscircle*[fillcolor=black](-0.6,7.5){0.02}
		\pscircle*[fillcolor=black](-0.6,7.3){0.02}
		\pscircle*[fillcolor=black](-0.6,7.7){0.02}
		\rput(6.35,6.47){\footnotesize $|\lambda_{2^N}\rangle$}
		\rput(6.5,6.8){\footnotesize $|\lambda_{2^N+1}\rangle$}
		\rput(6.5,7.13){\footnotesize $|\lambda_{2^N+2}\rangle$}
		\rput(6.65,7.9){\footnotesize $|\lambda_{2^{N+1}-1}\rangle$}
		\psline[linestyle=dashed](-0.5,6.3)(0.7,6.3)
		\psline[linecolor=red]{<->}(-0.2,1.5)(-0.2,6.3)
		\psline[linecolor=red]{<->}(1.4,2)(1.4,6.8)
		\psline[linestyle=dashed](-0.5,6.1)(0.7,6.1)
		\psline[linecolor=blue]{<->}(0.1,1.5)(0.1,6.1)
		\psline[linecolor=blue]{<->}(1.7,2.5)(1.7,7.1)
		\psline[linestyle=dashed](-0.5,5.9)(0.7,5.9)
		\psline[linecolor=green]{<->}(0.4,1.5)(0.4,5.9)
		\psline[linecolor=green]{<->}(2,3.6)(2,7.9)
		\pszigzag[coilarm=.2,linearc=.1,coilwidth=.4]{->}(4.5,7.1)(4.5,5)
		\pszigzag[coilarm=.2,linearc=.1,coilwidth=.4]{->}(3.8,6.8)(3.8,5)
		\pszigzag[coilarm=.2,linearc=.1,coilwidth=.4]{->}(3.1,6.5)(3.1,5)
		\pszigzag[coilarm=.2,linearc=.1,coilwidth=.4]{->}(5.2,7.9)(5.2,5)
	}
	\rput(0,0.2){
	\rput(0.3,3.5){$2 \chi/\hbar$}
	\pscurve{->}(0.8,3.6)(0.95,3.85)(1.2,3.9)
	\pscurve{->}(0.8,3.4)(1.1,3.1)(1.8,3)
	\pscurve{->}(0.8,3.5)(1.1,3.7)(1.5,3.7)
	}
	\rput(3.6,0){
	\rput(0.7,3.5){$2 \chi/\hbar$}	
	\pscurve{->}(0.2,3.6)(0.05,3.85)(-0.2,3.9)
	\pscurve{->}(0.2,3.4)(-0.1,3.1)(-0.8,3)
	\pscurve{->}(0.2,3.5)(-0.1,3.7)(-0.5,3.7)
	}
	\endpspicture
		\caption{Relevant level scheme for a the cooling of $N$ qubits into the ground state of $H_{\rm Int}$. The states $|\lambda_n \rangle$ with $0 \le n < 2^N$ are eigenstates of $H_{\rm Int}$. Moreover, $2^N-1$ laser fields couple $|\lambda_n \rangle$ with Rabi frequency $2 \chi/\hbar$ to $|\lambda_{2^N+n} \rangle$ with one atom in an excited state. The different colors indicate different laser frequencies. These are chosen such that $|\lambda_0 \rangle$ is the only state without resonant driving. For simplicity, the figure does not show the off-resonant driving of the other qubit states. All qubit states are possible reset states in case of the spontaneous emission of a photon.}\label{Nqubits}
	\endfigure

In the $N$-qubit case and in the presence of the cooling lasers, the Hamiltonian of the system is given by Eq.~(\ref{Hatoms2}) with $H_{\rm Free}$ as in Eq.~(\ref{HFree}), $H_{\rm Int}$ as in Eq.~(\ref{Hint2}), and $H_{\rm Laser}$ as in Eq.~(\ref{Hlaser}). As in the one-qubit case, it is useful to express this Hamiltonian as a function of the $|\lambda_n \rangle$-states introduced in Section \ref{SecCoolingDevice}. In analogy to Eq.~(\ref{Hint2}), $H$ can be written as
	\begin{eqnarray}\label{Hint3}
		H &=& \sum_{n=0}^{4^N-1} E_n \, |\lambda_n \rangle \langle \lambda_n| + \sum_{n=0}^{4^N-1} \sum_{m\neq n} \chi_{mn} \, |\lambda_m \rangle \langle \lambda_n| 
	\end{eqnarray}  
with
\begin{eqnarray}
E_n \equiv \langle \lambda_n | H |\lambda_n\rangle ~~ {\rm and} ~~ \chi_{m n} \equiv \langle \lambda_m | H |\lambda_n\rangle \, .
\end{eqnarray}
The $E_n$-terms in the Hamiltonian (\ref{Hint3}) are effective level shifts, while the $\chi_{m n}$-terms are
time-dependent since they correspond to laser-driven transitions. In the case of states with at least one atom in an excited state they also include the effect of the interaction Hamiltonian $H_{\rm Int}$. 

As mentioned before, the vectors $|\lambda_n \rangle$ with $n$ between $0$ to $2^N-1$ are eigenvectors of $H_{\rm Int}$. From this one can easily see that
\begin{eqnarray}
E_n= \hbar \omega + \lambda_n ~~ {\rm and} ~~ \chi_{mn} = 0 ~~ {\rm for} ~ 0 \le n,m < 2^N \, .
\end{eqnarray}
For simplicity we assume again that the laser Rabi frequencies $\Omega_j^{(i,k)}$ in Eq.~(\ref{Hlaser}) are all the same and given by $\Omega$. In analogy to Eq.~(\ref{sirene}), we moreover introduce states $|\lambda_m \rangle$ with one atom in $|e_0 \rangle$ or $|e_1 \rangle$ and $m$ between $2^N$ and $2^{N+1}-1$ as
	\begin{eqnarray}\label{lambdaex2}
		|\lambda_m \rangle &\equiv & \frac{1}{\sqrt{N}} \Bigg[\sum_{i=0}^N \sum_{j=0}^1 \, |e_j \rangle_{ii} \langle g_j |\Bigg] |\lambda_{m-2^N} \, \rangle \, .
	\end{eqnarray}
The energy $E_m$ of these states is given by
\begin{eqnarray}
E_m= \hbar \tilde \omega + \langle \lambda_m | H_{\rm Int} |\lambda_m \rangle \
\end{eqnarray}
and they are pairwise orthonnormal. Indeed they complete the basis which contains the first $2^N$ states $|\lambda_n \rangle$ further. From Eq.~(\ref{Hlaser}) we see that the $\chi_{mn}$ with $0 \le n < 2^N \le m < 2^{N+1}$ depend neither on $n$ nor $m$. For them we can hence assume $\chi_{mn} = \chi$ with 
\begin{eqnarray}
\chi (t) &\equiv & \sum_{k=1}^{2^N-1} {1 \over 2} \hbar \sqrt{N} \Omega \, {\rm e}^{{\rm i} \big( \tilde \omega - \omega -\Delta_k \big) t} \, .
\end{eqnarray}
The reason for this simplification is that the applied laser fields drive all transitions with the same Rabi frequency. 

The main transitions involved in the time evolution of the system are shown in Fig.~\ref{Nqubits}. In order to cool into the ground state $|\lambda_0\rangle$ it is sufficient to tune for example the frequency of laser $k$ in resonance with the $|\lambda_k \rangle$ - $|\lambda_{k+2^N}\rangle$  transition. This means, the frequency $\omega_k$ of laser $k$ should be chosen such that
\begin{eqnarray} 
\hbar \omega_k &=& E_{2^N+k} - E_k \, .
\end{eqnarray}
This implies $\hbar \Delta_k = \langle \lambda_{2^N+k} | H_{\rm Int} |\lambda_{2^N+k} \rangle  - \lambda_k $, since $\Delta_k = \tilde \omega - \omega - \omega_k$. Moreover, we notice that the detunings of the applied laser fields with respect to $|\lambda_0\rangle$ are given by 
\begin{eqnarray} 
E_{2^N} - E_0 -  (E_{2^N+k} - E_k) &=& \langle \lambda_{2^N} | H_{\rm Int} |\lambda_{2^N} \rangle - \langle \lambda_{2^N+k} | H_{\rm Int} |\lambda_{2^N+k} \rangle \nonumber \\ 
&& + \lambda_k - \lambda_0  \, .
\end{eqnarray}
If $k$ ranges from $1$ to $2^N-1$, then the ground state $|\lambda_0\rangle$ is the only state which does not experience resonant driving.

The comparison with laser sideband cooling suggests that population accumulates in $|\lambda_0\rangle$ as long as the effective driving experienced by this state is much weaker than the driving experienced by the $|\lambda_k \rangle$ states with $0 < k < 2^N$. More concretely, in analogy to Eq.~(\ref{comparison}), the system parameters should be chosen such that
\begin{eqnarray} \label{cond}
{1 \over \hbar} \, |E_{2^N} - E_0 -  (E_{2^N+k} - E_k)| & \gg & \sqrt{N} \Omega ~~ {\rm and } ~~ \Gamma 
\end{eqnarray}
for all $k$. Notice that this condition poses an upper bound on the achievable cooling rate for a given interaction Hamiltonian $H_{\rm Int}$. However, it should be possible to speed up the cooling process and to obtain nevertheless relatively high fidelities by slowly decreasing the Rabi frequency $\Omega$ in time. While $\Omega $ becomes smaller, the fidelity of the final state can become arbitrarily close to unity. Once populated, we expect that the atoms remain much longer in $|\lambda_0 \rangle$ than in any other qubit state. 

\section{Concrete examples} \label{sec4}

In this section, we explicitly calculate the achievable fidelities and the corresponding cooling rates for the one-qubit case with an arbitrary interaction Hamiltonian $H_{\rm Int}$. Afterwards, we present numerical results for the two-qubit case for the concrete example of a spin-spin Heisenberg interaction and describe the cooling of the atoms into a maximally entangled state.

\subsection{1-qubit case}

In the one-qubit case, the fidelity and the cooling rate of the proposed state preparation scheme can be obtained easily by calculating the  stationary state of the system analytically. Using the master equation (\ref{mastereq}) for the time-independent Hamiltonian $H_{\rm I}$ in Eq.~(\ref{U0HlaserU0}) and setting $\dot \rho = 0$, we find that the matrix elements of the stationary state with respect to the basis $\{|\lambda_0\rangle, |\lambda_1\rangle, |\lambda_2\rangle, |\lambda_3\rangle\}$ are given by
	\begin{eqnarray}\label{stationarystate}
		&& \rho_{00} = \frac{4 \Gamma^2 + 4 \Delta_\lambda^2 +\Omega^2}{4(2\Gamma^2 + \Delta_\lambda^2 +\Omega^2)} \, , \nonumber \\
		&& \rho_{11} = \frac{ 4 \Gamma^2 +\Omega^2}{4(2\Gamma^2 + \Delta_\lambda^2 +\Omega^2)} \, , \nonumber \\
		&& \rho_{22} = \rho_{33} = \frac{\Omega^2}{4(2\Gamma^2 + \Delta_\lambda^2 +\Omega^2)} \, , \nonumber \\
		&& \rho_{02} = \rho_{20}^* = \frac{- \Delta_\lambda \Omega + {\rm i} \Gamma \Omega}{2(2\Gamma^2 + \Delta_\lambda^2 +\Omega^2)} \, , \nonumber \\
		&& \rho_{13} = \rho_{31}^* = \frac{- {\rm i} \Gamma \Omega}{2(2\Gamma^2 + \Delta_\lambda^2 +\Omega^2)} 
	\end{eqnarray}
and
	\begin{eqnarray}
		&& \rho_{01} = \rho_{03} =\rho_{12} = \rho_{23} =  \rho_{10} = \rho_{30} =\rho_{21} = \rho_{32}  = 0 
	\end{eqnarray}
with $\Delta_\lambda$ defined in Eq.~(\ref{Delta}). The fidelity $F$ for the preparation of $|\lambda_0 \rangle$ is therefore given by $\rho_{00}$, since the state preparation is complete once the system reached its stationary state. This means
	\begin{equation}
		F=1 - \frac{4\Gamma^2 + 3 \Omega^2}{4(\Delta_\lambda^2 + 2 \Gamma^2 + \Omega^2)} \, .
	\end{equation} 
As shown in Fig.~\ref{FvsDelta}, this fidelity can be arbitrarily close to unity. As suggested by Eq.~(\ref{cond}), high fidelities are obtained when $\Delta_\lambda$ is  much larger than $\Omega$ and $\Gamma$.

	\figure\quad\quad\quad\quad\quad
	\subfigure[]{\includegraphics[width=.41 \columnwidth]{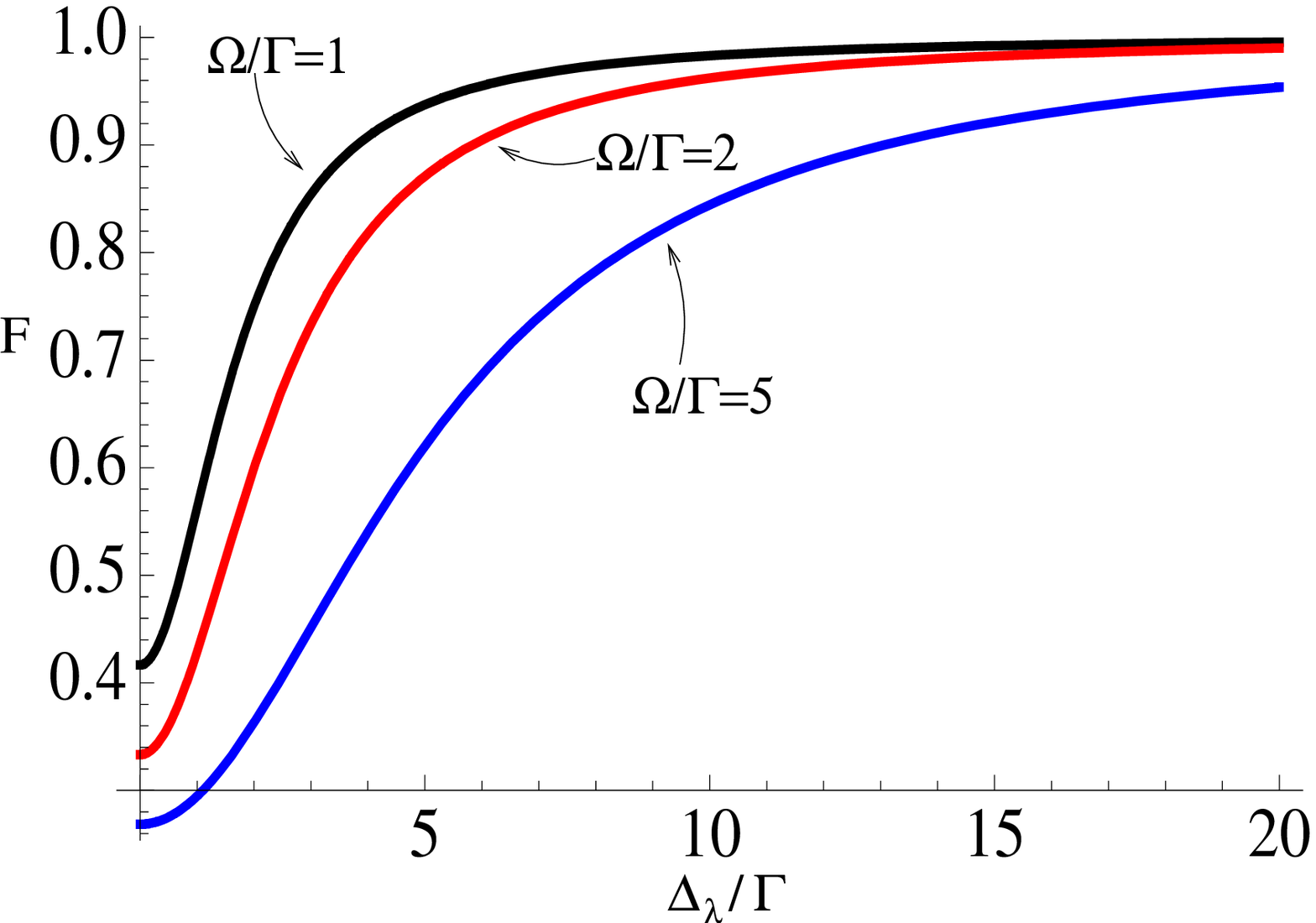}}\quad
	\subfigure[]{\includegraphics[width=.41 \columnwidth]{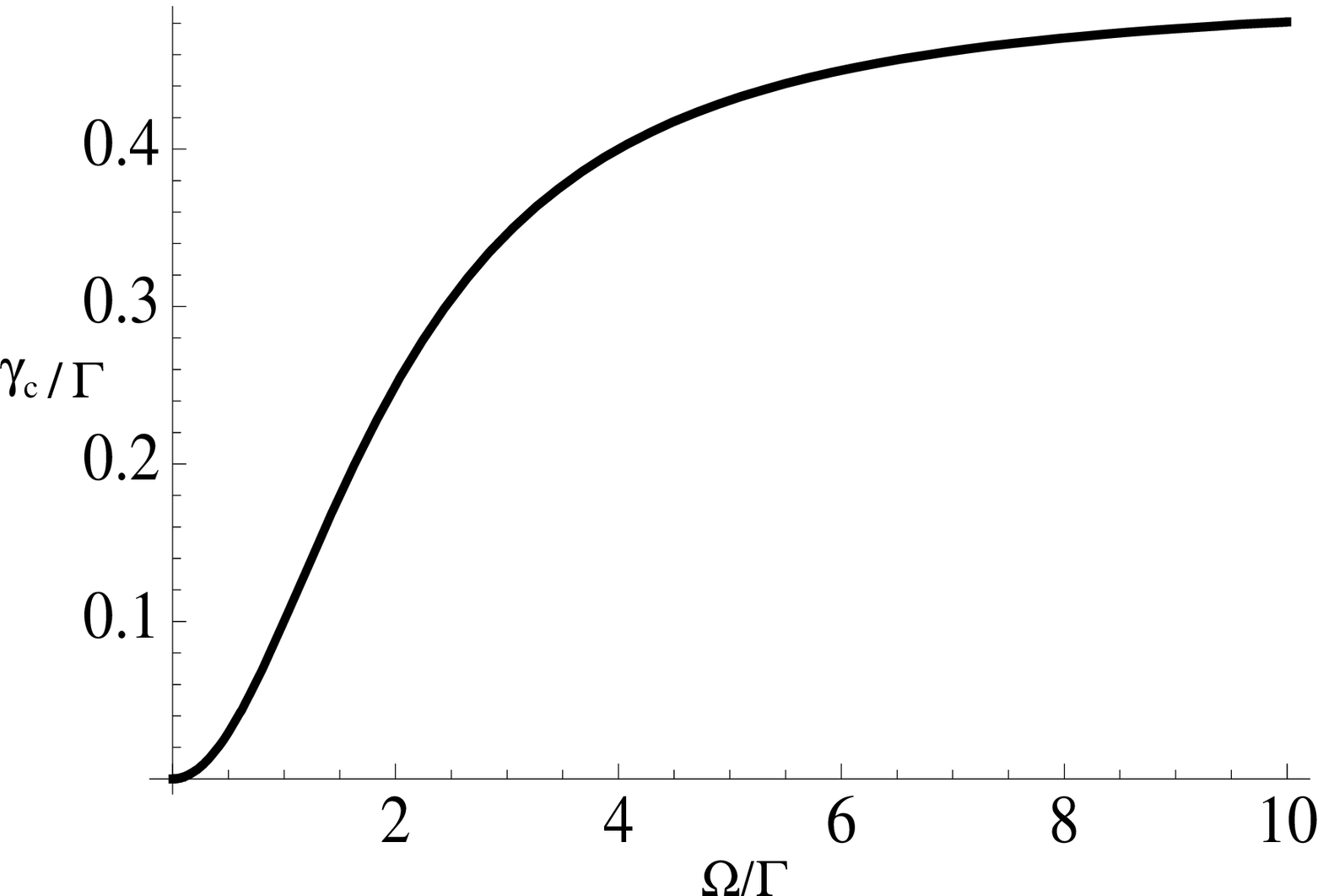}}
		\caption{(a) Fidelity for the preparation of the ground state $|\lambda_0\rangle$ as a function of the effective detuning $\Delta_\lambda$ for different Rabi frequencies $\Omega$.  (b) Cooling rate $\gamma_{\rm c}$ as a function of $\Omega / \Gamma$ for large detunings $\Delta_\lambda$ obtained from Eq.~(\ref{coolingrate2}) .} \label{FvsDelta}
	\endfigure

The cooling rate $\gamma_{\rm c}$ gives an indication for how long it takes to realise the above fidelity. To calculate it for the proposed state preparation scheme, we note the probability flux conservation
	\begin{eqnarray} \label{flux}
		\gamma_{\rm h} \, \rho_{00} = \gamma_{\rm c} \, (1 - \rho_{00}) \, .
	\end{eqnarray}
Here $\gamma_{\rm h}$ is the heating rate, {\em i.e.} the rate with which a system prepared in $|\lambda_0 \rangle$ leaves the target state. To obtain $\gamma_{\rm h}$ we note that leaving $|\lambda_0 \rangle$ is only possible via the accumulation of a small amount of population in $|\lambda_2 \rangle$ due to non-resonant laser driving followed by the spontaneous emission of a photon with decay rate $\Gamma/2$ into $|\lambda_1 \rangle$. Hence
	\begin{eqnarray}
		\gamma_{\rm h} = {1 \over 2} \Gamma \, \rho_{22} \, .
	\end{eqnarray}
Since the laser driving of the $|\lambda_0 \rangle$ - $|\lambda_2 \rangle$ transition is strongly detuned, this population is more or less constant in time, even before the system reaches its steady state. Approximating $\rho_{22}$ by its steady state value in Eq.~(\ref{stationarystate}) and using Eq.~(\ref{flux}), we finally obtain the cooling rate
	\begin{equation}\label{coolingrate}
		\gamma_{\rm c} = \frac{\Gamma \Omega^2 ( 4 \Delta_\lambda^2 + 4 \Gamma^2 + \Omega^2 )}{8(\Delta_\lambda^2 + 2 \Gamma^2  + \Omega^2)(4 \Gamma^2  +3 \Omega^2)} \, .
	\end{equation}
For large detunings $\Delta_\lambda $, this equation simplifies to
	\begin{equation}\label{coolingrate2}
		\gamma_{\rm c} = \frac{\Gamma \Omega^2}{2(4 \Gamma^2  +3 \Omega^2)} \, .
	\end{equation}
As illustrated in Figure \ref{FvsDelta}, this rate no longer depends on $\Delta_\lambda$. Maximum cooling rates mainly require an as large as possible value for $\Omega$ without violating condition (\ref{cond}). Moreover, $\Gamma $ should not be much larger than $\Omega$. Otherwise, the interaction with the environment results in continuous measurements which freeze the atom as predicted by the quantum Zeno effect \cite{Misra} and make it impossible to reach the target state.

\subsection{2-qubit case} \label{viola}

	\figure
	\quad\quad\quad\quad
	\includegraphics[width=.6 \columnwidth]{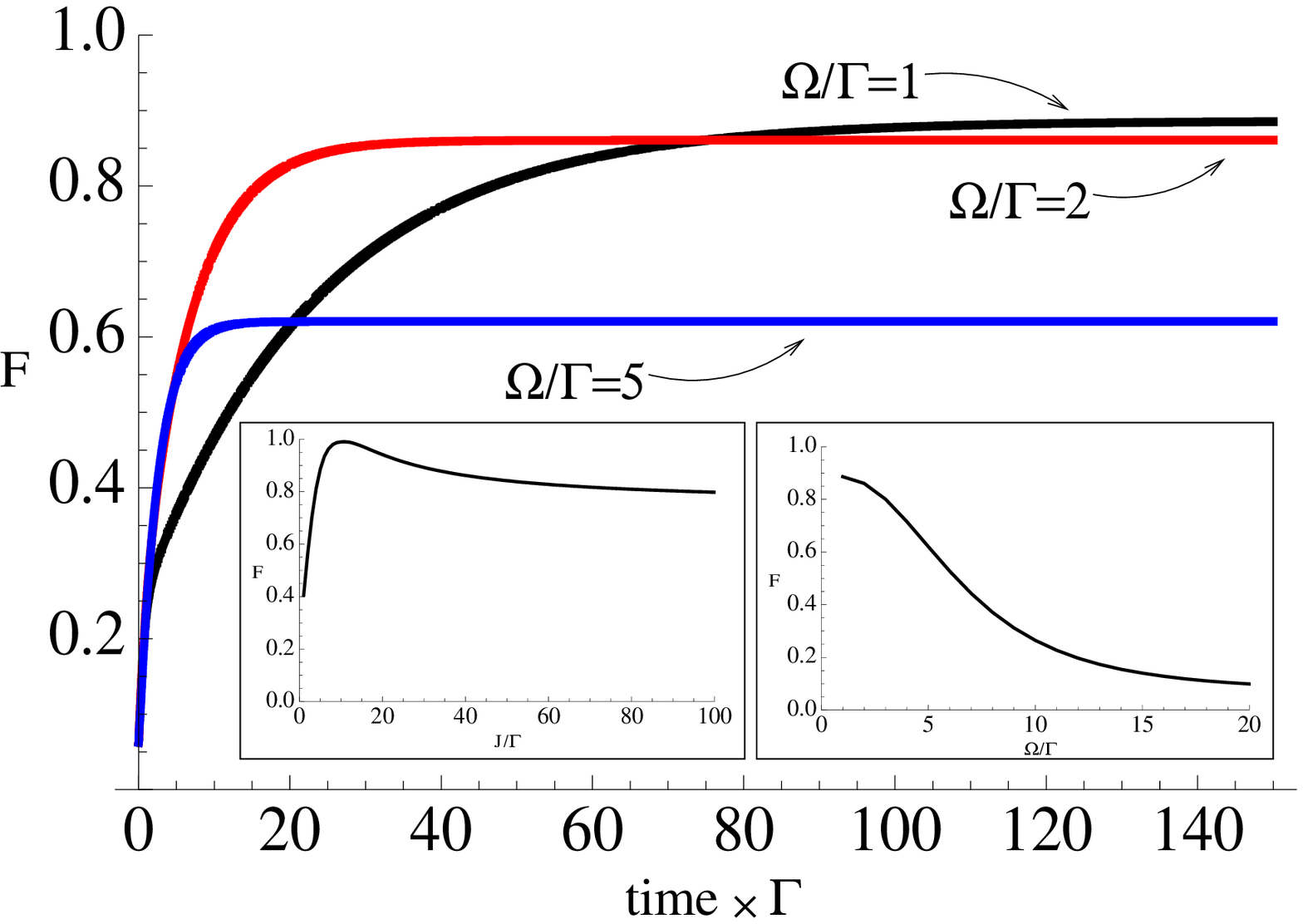}\\
		\caption{Fidelity $F$ for the preparation of the maximally entangled state of two qubits as a function of  time $t$ for $J=5 \, \Gamma$ and for different $\Omega$'s. The small figures show $F$ as a function of the coupling constant $J/\Gamma$ for $\Omega =\Gamma$ and  as a function of the Rabi frequency $\Omega/\Gamma$ for $J =5 \, \Gamma $.}\label{F2Qubits}
	\endfigure

In this section we discuss  the preparation of two qubits in a maximally entangled state. As a concrete example, we consider a particular case of the spin-spin Heisenberg Hamiltonian and assume
	\begin{equation}\label{spinspinH}
		H_{\rm Int} = \hbar J \,  \vec{\sigma}_1 \cdot \vec{\sigma}_2 \ ,
	\end{equation}
where $J$ is a coupling constant and  $\vec{\sigma}_i \equiv (\sigma_i^x, \sigma_i^y, \sigma_i^z)$. This means, we treat the two ground states of each atom as a pseudo-spin described by the Pauli matrices $(\sigma^x, \sigma^y, \sigma^z)$. In terms of its eigenvectors $|\lambda_n\rangle$ and eigenvalues $\lambda_n$, the Hamiltonian (\ref{spinspinH}) can be written as
	\begin{equation}\label{spinspinH2}
		H_{\rm Int} = -3 \hbar J \, |\lambda_0\rangle \langle \lambda_0| +  \sum_{n=1}^{3} \hbar J \, |\lambda_n \rangle\langle \lambda_n| 
	\end{equation}
with 
 \begin{eqnarray}
 |\lambda_0\rangle &=& (|01\rangle -|10\rangle)/\sqrt2 \, , \nonumber\\
  |\lambda_1\rangle &=& (|01\rangle +|10\rangle)/\sqrt2 \, , \nonumber\\
   |\lambda_2\rangle &=&|00\rangle \, , \nonumber\\
    |\lambda_3\rangle &=&|11\rangle \, ,
 \end{eqnarray}
and $\lambda_0 = -3\hbar J$, while $\lambda_1=\lambda_2=\lambda_3= \hbar J$.  In the following, we calculate the stationary state of the system in the presence of the cooling lasers in order to determine the fidelity of the state preparation.

As in the one-qubit case in Eq.~(\ref{CondParam}) we assume that the laser Rabi frequencies and the decay rates are for all transitions the same. In analogy to Eq.~(\ref{sirene}) we introduce the states 
	\begin{eqnarray} \label{sirene2}
		|\lambda_{n+4} \rangle &\equiv & \frac{1}{\sqrt{2}} \left[ \sum_{i=1,2}\sum_{j = 0,1}|e_j \rangle \langle g_j | \right] | \lambda_n \rangle
	\end{eqnarray}
with one atom in $|e_0 \rangle$ or $|e_1 \rangle$. For simplicity and since it is anyway small, we neglect population in the states with both atoms excited. In this case, the time evolution of the system remains restricted onto an eight-dimensional subspace for which the vectors $|\lambda_n\rangle$ with $n$ between 0 and 7 form a complete basis. In the interaction picture with respect to $H_0 = H_{\rm Free} - \hbar J \sum_{i=1,2} \sum_{j=0,1} |e_j \rangle_{ii} \langle e_j|$, the system Hamiltonian including the laser driving can now be written as
	\begin{eqnarray}
		H_{\rm I} &=& \sum_{n=0}^{3} \frac{1}{2} \, \hbar \sqrt{2} \Omega \, |\lambda_n \rangle \langle \lambda_{n+4}|  + {\rm H.c.} +  \hbar (\lambda_n - \Delta) \, |\lambda_n \rangle \langle \lambda_n| \, .
	\end{eqnarray}
	When the laser detuning $\Delta$ equals $\hbar J$, then the states $|\lambda_1\rangle$, $|\lambda_2\rangle$ and $|\lambda_3\rangle$ experience a resonant driving, while $|\lambda_0\rangle$ is off-resonance. In the concrete example considered here, the time evolution of the system is analog to the one-qubit case. Under the condition of sufficiently small Rabi frequencies $\Omega$ and decay rates $\Gamma$, it is possible to achieve fidelities well above $90 \%$, as shown in \figurename~\ref{F2Qubits}.

\section{Conclusions} \label{sec6}

This paper discusses how to cool a system of atomic qubits into the ground state $|\lambda_0 \rangle$ of an applied interaction Hamiltonian $H_{\rm Int}$. Our cooling device consists of laser fields which couple the qubit states $|g_0 \rangle =|0 \rangle$ and $|g_1 \rangle = |1 \rangle$ to auxiliary atomic states $|e_0 \rangle$ and $|e_1 \rangle$ with non-zero spontaneous decay rates. Laser frequencies should be chosen such that the driving of $|\lambda_0 \rangle$ is off-resonant, while all other qubit states experience resonant laser driving. Once spontaneous emission results in the population of $|\lambda_0 \rangle$, the system remains there to a very good approximation for a wide range of experimental parameters and independent of the initial state of the system. For simplicity, we assume degenerate qubit states and degenerate excited atomic states and equal laser Rabi frequencies and decay rates for all possible atomic transitions. In this way, it is easy to change from one coordinate system into another and complex notation has been avoided. However, the application of the proposed cooling mechanism to a large variety of atomic systems is straightforward. 

Preparing the ground state of $N$ qubits of an arbitrary interaction Hamiltonian $H_{\rm Int}$ can require adjusting up to $2^N-1$ different laser frequencies. However, as we have seen in Section \ref{viola}, this number is significantly lower in case of degeneracies. In concrete situations, it is even possible to realise the state preparation with the help of optimised laser pulse sequences \cite{sophie} which naturally contain a wide range of frequencies. As in laser sideband cooling, high fidelities require that the detunings seen by $|\lambda_0 \rangle$ are much larger than the spontaneous decay rate of $|e_0 \rangle$ and $|e_1 \rangle$ and the effective Rabi frequencies which couple the qubit states to states with one atom in $|e_0 \rangle$ or $|e_1 \rangle$ (cf.~Eq.~(\ref{cond})). Efficient cooling therefore requires an interaction Hamiltonian $H_{\rm Int}$ for which the difference between the energy of the $|\lambda_0 \rangle$ - $|\lambda_{2^N} \rangle$ transition and the energy of the $|\lambda_n \rangle$ - $|\lambda_{n+2^N} \rangle$ transitions with $n>0$ is sufficiently large. This means, similar to adiabatic quantum computation \cite{review}, the proposed cooling scheme requires a relatively well distinguished ground state.

Our discussion is in good agreement with a detailed analysis for examples with one and two qubits. In the one-qubit case, it is possible to calculate the achievable cooling rates and fidelities analytically by simply deferring them from the stationary state of the system. While the cooling rate increases in general with the laser Rabi frequency $\Omega$ (cf.~Fig.~\ref{FvsDelta}(b)), high fidelities are easier to obtain for relatively small $\Omega$'s (cf.~Fig.~\ref{FvsDelta}(a)). One should therefore decrease the laser intensity during the cooling process in order to obtain high speed and high precision. In the two-qubit case, we consider a Heisenberg interaction Hamiltonian $H_{\rm Int}$ with three degenerate eigenstates and a non-degenerate ground state $|\lambda_0 \rangle$. 

Like most measurement-based state preparation schemes, the cooling process is expected to be very robust against parameter fluctuations. Nevertheless, there is no need to register certain measurement outcomes. An interesting question is the efficiency of the proposed cooling scheme for a large number of qubits.  Since the interaction Hamiltonian $H_{\rm Int}$ acts not only on the qubit states but also on states with one atom excited, the energy gap on the left hand side of Eq.~(\ref{cond}) decreases rapidly as $1/N$ for $M$-local Hamiltonians with $M \ll N$. To increase the efficiency of the proposed cooling mechanism for large numbers of qubits, one could create entanglement for example by initially cooling only separate cells of a finite size. This entanglement can then be distributed via controlled interactions between neighboring cells. \\

\noindent {\em Acknowledgement.} The authors thank Gavin Brennen and Alex Monras-Blasi for stimulating and very helpful discussions. A. B. acknowledges a James Ellis University Research Fellowship from the Royal Society and the GCHQ. This work was supported in part by the UK Research Council  EPSRC, the EU Integrated Project SCALA, and the EU Research and Training Network EMALI.

\end{document}